\documentclass{conm-p-l}%
\usepackage{amsfonts}
\usepackage{amsmath}
\usepackage{amssymb}
\usepackage{graphicx}%
\providecommand{\U}[1]{\protect\rule{.1in}{.1in}}
\theoremstyle{definition}

\theoremstyle{remark}

\numberwithin{equation}{section}
\theoremstyle{plain}

\copyrightinfo{2001}{enter name of copyright holder}
\ifx\pdfoutput\relax\let\pdfoutput=\undefined\fi
\newcount\msipdfoutput
\ifx\pdfoutput\undefined\else
\ifcase\pdfoutput\else
\ifx\paperwidth\undefined\else
\ifdim\paperheight=0pt\relax\else\pdfpageheight\paperheight\fi
\ifdim\paperwidth=0pt\relax\else\pdfpagewidth\paperwidth\fi
\fi\fi\fi
\begin{document}
\title[Trilogy of ANW]{Whitehead's Trilogy and the Curavature of Spacetime}
\author{A. John Coleman}
\address{Department of Mathematics and Statistics, Queens University, Kingston, ON Canada.\\
\"{y} }
\email{colemana@post.queensu.ca}
\date{April 20, 2007}

\begin{abstract}
We outline the basic ideas of Whitehead's unified theory of Gravitation and
Electromagnetism; summarize the evidence that Spacetime is curved and conclude
that it is not.

\end{abstract}
\maketitle

\section{$\bigskip$}

\section{\bigskip Whitehead's Trilogy and the Curvature of Spacetime}

\begin{center}
A John Coleman, Department of Mathematics and Statisitcs, Queens Univerisity,
Kingston, Ontario, Canada.

\textit{colemana@post.queensu.ca}

\bigskip
\end{center}

\subsection{I. \ Introduction}

In the three years 1919-1922, Alfred North Whitehead(ANW) published three
profound books$^{1}$ unifying Gravitation and Electromagnetism. This Trilogy,
in my opinion, contains the most important contribution to fundamental
physical theory by any one individual since Newton. As far as I have
discovered, it has hardly been noticed in the past 40 years by mainstream
Physicists who have unfortunately allowed Whitehead to be pre-empted by
philosophers even though he was professionally occupied as a Mathematical
Physicist until he was 63!

The\ first two volumes of the Trilogy are introductory to the\ third entitled
\textit{Principle of} \textit{Relativity with applications to Physical
Science(PRel).} $\mathit{\ }$In this, ANW shows\ that there is a fundamental
logical flaw in the General Theory of Relativity(GTR)$^{2}$ such that \ in its
current form, any physical theory based on GTR is inevitably flawed. \ He
insists that experiment and logic are the ultimate tests of a physical theory
and that there is no \textit{valid} evidence to indicate that Spacetime (ST)
is curved. This claim cannot be lightly dismissed since it was Whitehead who
with his former "pupil", Bertrand Russell, wrote the famous three-volume
\textit{Principia Mathematica }in Symbolic Logic establishing themselves as
the two greatest Logicians of the past 200 years. \ 

\ Since I taught GTR enthusiastically at the University of Toronto for 8
years, it is my duty to admit my error and explain how I and most Mathematical
Physicists were misled by Eddington. \ 

We shall see that the essential problem is that of \textit{definition of
units. }It was Eddington who got us into the difficulty in 1920. In my
Concluding Observations I shall point to an idea from his \textit{Fundammental
Theory }which suggests that by 1947 he realized that defining units is an
important non-trivial difficulty.

The Trilogy is perhaps the high-point in a series of nine$^{3}$ major works
ANW published between 1905 and 1929 providing us with insight into his attempt
to define concepts and words with which to free himself and us from the
brain-washing in which he lived, and we still live, about the meaning of
reality, space, time and matter, ideas for example which we inherited \ about
substance from Aristotle and about\ space or time from Newton. \ His 1926-27
Gifford Lectures, sub-titled \textit{An Essay in Cosmology} is perhaps the
only Metaphysical framework proposed by a scientist with an understanding of
the Special and General Theories of Relatvity and of Planck's Quantum Theory. \ 

\bigskip

My object in the present paper is to:

\qquad- introduce physicists to the life and work of Whitehead, Sec. II;

\qquad- outline the content of \textit{The Principle of Relativity }(PRel),
Sec. III;

\qquad- present ANW's basic criticism of GTR, which , to my mind, has never
been rebutted or properly appreciated, even by GTR specialists, Sec. IV;

\qquad- propose a significant relevant research problem, Secs.VI \& VIII;

\qquad- summarize some of the rather radical implications of the acceptance of
ANW's observation.

\bigskip

\subsection{II. LIfe and Work\ (1861-1947)}

His career is best characterized as that of a Mathematical Physicist$^{4}$
seeking total understanding of his own physical nature and its place in the
ongoing process which we experience as human beings and call the Universe.

Alfred was born on Febuary 15, 1861 in a family of Anglican clergy, school
masters and local Administrators. His father was a priest, a school master and
an honorary Canon of Canterbury Cathedral which is the spiritual centre of the
Anglican Communion throughout the world.

In 1876, Whitehead entered Sherborne School which, according to tradition, had
been attended by King Alfred the Great and of which Whitehead becane Senior
Prefect responsible for behaviour and discipline. He won a scholarship to
Trinity College, Cambridge - majored in Mathematics and graduated in 1884 as
Fourth Wrangler in the demanding\ Tripos Examination. He was soon named a
Fellow of Trinity College partly as a result of an Essay on Maxwell's Theory
which was still quite novel. \ As a Fellow his basic task was the preparation
of College students to master Mathematical Analysis, Analytical Dynamics of
Lagrange and Hamilton together with Newton's theory of Gravitation and
Maxwell's theory of Electrodynamics. The latter was his favourite topic for
university Lectures.

Among his early mathematical writings were papers on approximating the Motion
of viscous fluids, Group theory, and Geodesic Geometry. A major book entitled
\textit{Universal Algebra }(1898) made him the first winner of the
international Lobatchevsky Award. This book contained the first clear
exposition in English of Grassmann's algebra which now plays a key role \ in
theoretical physics. \ It is really the language of the QM of fermions and of
exterior differential forms.

From 1905 to 1912, Whitehead's big project, together with Bertrand Russell,
was writing the three volume treatise, \textit{Principia Mathematica,}\ on the
foundations of Mathematics.

He decided in 1912 to move from Cambridge and was soon established in the
University of London as Professor of Applied Mathematics at the
\textit{Imperial College of Science and Technology} which vies with MIT in the
USA, and ETH\ \ in Switzerland as among the truly great Engineering schools.\ 

From there, when he was 63, he was invited to become a Professor of Philosophy
in Harvard. He made an enormous impact on philosophical thought epitomized in
his \textit{PROCESS and REALITY} which soon became the basis of a totally new
important School of philosophy generally referred to as \textit{Process
Philosophy }for which "process" is an ultimate category. \ 

For physicists this implies that we must learn to think of the basic
constituents of physical reality as "events" not as bits of red or black
Aristotelian "substance". To imitate a famous phrase$^{5}$ of Reinhold
Niebhur, this will require a "trans-imagining of our imagination"! I am far
from accomplishing this even though I have known for more than \ sixty years
that this is required of the physics community. \ I was 20 years old$^{6}$
when, accidentally, I first tackled PRel.

From 1924 umtil his death, Whitehead wrote prolifically, conducted Graduaate
Seminars at Harvard and lectured widely /throughout the USA applying his deep
intelligence to all the problems of mankind. For example, I found his books on
\textit{The Function of Reason }\ and \ \textit{Adventures of Ideas
}especially interesting, even exciting. For more detailed information about
ANW's life and writings as Mathematical Physicist and as Philosopher I advise
the reader to consult the definitive biography by Lowe$^{7}.$

\subsection{}

\subsection{III. \ Whitehead's \textit{Principle of Relativity}.(PRel)}

\textit{Throughout this Article I use the abbreviation "PRel" to denote
Whitehead's book}$^{1},$\textit{The Principle of Relativity with Applications
to Physical Science, and PNK \ and CN for the other}$^{1}$\textit{ two books
in the Trilogy.}

There \ was so little interest in PRel that the Cambridge University Press
never reprinted the 1922 First Edition. Fortunately after the Copyright
lapsed, Dover and Phoenix,\ Publishers,\ recently made the original text
available. There is now even a paperback Edition so anyone interested has easy
access to it. \ I entertain three reasons why Physicists have dismissed PRel:
\ (1) the Einstein Euphoria$^{8}$which has been thoroughly documented, (2) the
disdain$^{9}$ that Physicists had for philosophy until quite recently, and (3)
admittedly, PRel\ is not an easy read. \bigskip

A paper$^{10}$ in\textit{\ Latex }to which I refer as $\left(  JLS\right)  $,
contains a brief Introduction and an Appendix on the Solar Limb Effect by
myself, but, chiefly, three lectures by \textit{John Leighton Synge }which
reproduce in current notation the contents of a mathematical exposition of
those aspects of PRel which are pertinent to the claims by Einstein to predict
the Advance of the Perihelion of Mercury and the Bending of star-light passing
close to the Sun.

\textit{Unfortunately Synge replaces ANW's notation }$J_{ij}$ \textit{for
Whitehead's gravitational impetus by }$g_{ij}$ causing great confusion for any
reader familiar with current expositions of GTR. He also , reveals, almost
proudly, that he did not read Part I seriously. He therefore missed the main
point of PRel.

\bigskip

\ The book is divided into three Parts:

I. \ \ General Principles, Chaps. 1-4

II. Physical Applications, Chaps. 5-17

III. Elementary Theory of Tensors, Chaps.\ 18-24

As Whitehead explained in his Preface, Part I assembles two formal lectures
which he gave in Bryn Mawr and the Royal Society of Edinburgh together with
material to students of The Imperial College. I found that, as he feared, this
and haste did cause some incoherence and lack of desirable transition material
at certain points.

Whitehead sent PRel to the Publisher near the beginning of 1922 so knew
perfectly well that the London and the New York Times had already elevated
Einstein to the status of the untouchable Icon he became. So, to justify his
temerity in criticising GTR in PRel and CN, he praised Einstein to he skies
adding the caveat: \ "The worst hommage we can pay to genius is to accept
uncritically the formulations of truths which we owe to it" - PRel p.88.

To no avail. The Einstein Euphoria swept him aside \ - like a man in a canoe
defying the tsunami which devastated S-E Asia! The fact pointed out by
Whitehead that there is an obvious LOGICAL flaw in GTR has, as far as I have
discovered, never been directly confronted in the Physics literature -
probably, not even noticed! \ However, it has been recently analyzed by two
philosophers$^{11}$.

In Part I of PRel, ANW announced \textit{four} possible self-consistent
unified theories of Gravitation and Electromagnetism each of which satisfies
the basic principles which guided Einstein to GTR, but assumes that ST is flat
Minkowskian. At least two, (i) and (iv), predict the observed values for the
Advance of the perihelion of Mercury and the Bending of Light in the field of
the sun. It was on the basis of these two results that in 1920 Eddington had
proclaimed GTR as correct.

I found that Part I contained some of Whitehead's most challenging writing
because \ in it he explicitly defines ideas which go beyond our traditional
physical concepts and presupposes that the reader has intimate understanding
of PNK and CN$^{1}$.

On PRel p. 86/7, he defines four plausisble distinct theories of Gravitation,
invariant with respect to Lorentz Transformations, each of which assumes that
ST is Minkowskian but differ by what he calls Law (i), (ii), (iii), or (iv).
\ In each case the "Law" specifies how the "gravitational \textit{IMPETUS}" is
defined by a quadratic form $J_{ij}dx^{i}dx^{j}.$\ The first three Laws
involve solving a coupled system of ten second\ order partial differential
equations. They were obtained by a relatively simple variation on Einstein's
procedure for obtaining the Field Equations of GTR.

In all of them, ST is Minkowskian with metric determned by a second order
covariant Tensor, $\omega_{ij},$for which orthogonal cartesian coordinates may
be chosen such that $\omega_{ij}=\omega_{i}^{2}\delta_{ij},$with $\omega
_{i}^{2}=1$ for $i=1,2,or3$ and $-1$ for $i=4.$ \ In other words, the familiar
metric for Minkowsky Spacetime.

These four theories of gravity together with his formula for the interaction
of the gravitational and elecromagnetic fields constitute what I am calling
his four possible Unified Field Theories. \ Theory (i) is closest to GTR while
assuming that space is flat.

Of course, GTR also has a metric defined by a second order covariant tensor
which is almost always denoted by $g_{ij}.$ \ According to ANW, Einstein's
crucial error is to identify $g_{ij}$ with $J_{ij}.$ As he says on PRel, p. 83
"By identifying the potential mass impetus of a kinematic element with a
spatio-temporal measurement Einstein, in my opinion, leaves the whole
antecedent theory of measurement in confusion when it is confronted with the
actual conditions of our perceptual knowledge."

For Einstein, the key functions, $g_{ij},$ are defined as the solution of 10
PDE's with initial conditions determined by the circumambient massive
particles and therefore, the whole massive Universe!. \ According to Whitehead
this is the crucial point at which Einstein introduces a \textit{vicious
circle}\ into the basis of \ GTR which is discussed in Section V.

ANW is frequently difficult to understand because he takes seriously the
changes in our classical mode of thinking about reality forced by the Special
Theory of Relativity. There is not space here to summarize this 24-year-long
development which began in 1905 with a paper \textit{On Mathematical Concepts
of the Material World, }followed in 1916 by \textit{Space,Time,and Relativity,
}in 1919-22 by the Trilogy \textit{PNK, CN, PR }and culminated in 1929 by his
famous metaphysical treatise, \textit{Process and Reality - an Essay in
Cosmology. }In these he introduced several new concepts and new words. But his
mind never rested so the exact meaning of some words changed, causing trouble
for us who struggle to understand his basic thinking!

A key concept in Whitehead's physics is \textit{IMPETUS. }This has a role in
Whiteheadian Dynamics analogous to "action" in classical dynamics. If we
denote "action" by A then classically we would think of the Principle of Least
Action as requiring that a particle P which goes from a point E to another
point F\ follows the path which minimizes the integral of dA along its path.
\ In my old unreformed mode, I would imagine the particle as a minute very
solid ball, probably black (but perhaps red, depending on my mood!) zooming
along a Feynmann diagram. Classically, I think of the Action as a property of
the \textit{instantaneous} little particle. For ANW only events have reality
so potential Impetus is ascribable to finite portions of the path of my
particle, so $\sqrt[2]{dJ^{2}}$ to the \textit{interval} (dX$^{i}).$Thus, not
to the instaneous particle as such but to portions of its history.

\textit{Impetus, I, }is defined mplicitly in Part I on p. 79\ of PRel in
equation $\left(  9\right)  $ which reads%

\[
dI=M\sqrt[2]{dJ^{2}}+c^{-1}EdF
\]
where \ $\sqrt[2]{dJ^{2}}$ is the potential gravitational mass impetus and
$dF$ is the potential electromagnetic impetus for the interval $dX,\left(
X,X+dX\right)  $. Thus impetus involved the forces known to physicists in 1922.

\ \ In the first of the Trilogy, on the \textit{Principles of Natural
Knowledge(PNK), 1919 ANW }pursued his attempt to find words and enunciate new
concepts with which to incorporate into our thinking the implications of
Special Relativity and Quantum Theory as they gradually emerged and were
modified during the period 1890 - 1924. This is a very difficult task since
both he and we have unconsciously been "brain-washed" by Aristotelian ideas of
`substance' and Newton's view of the relation of `time' and `space'. His
attempt began in 1905$^{3}$ and culminated in 1927 with his great metaphysical
treatise. So persons like myself or Synge, Schild and Will, and many
philosophers who think we have understood Whitehead's theory are likely wrong
since we have aimed at a moving target! \ 

It is often alleged that it is difficult to understand his writing because ANW
is confused and his style is bad. \ In fact, I find that his style is normally
clear\ even dramatic as is illustrated by his description of the 1919 Meeting
of the Royal Society at which Eddington announced the results of the
Observation, for which he was responsible, of the bending of light by the sun.
This was proclaimed by Eddington and accepted, by the London and New York
Times, as "proof " that Einstein's GTR was correct.\ 

ANW had no hesitation in agreeing that Eddington's Observation showed that
Newton's theory of gravitation was inadequate\ but he did not accept GTR as
finally correct because he quickly noted a basic Logical error and also, as is
clear from many assertions particularly in his discussion with Lucien
Price$^{12},$ he did not think that human beings, \textit{including himself},
are capable of formulating in human language ultimate truth. \textit{\ This
may be his basic difference from Einstein.}

\ Hitherto I have seen no\ clear written evidence of \textit{when} he realized
the error in GTR but from Chapter 8 of CN and my estimate of the time needed
to invent and calculaate the consequences of his Law(iv) as set forth in PRel,
I consider sometime in\ 1917 as most likely. Since on his instructions all his
personal files were destroyed upon his death we shall probably never know the
exact date but I shall now speculate how he was led to Theory (iv).

As soon as he saw the text of Einstein's 1915 paper, ANW saw he logical flaw
\ in it and also thought that the Einstein's prediction of the observed value
for the Advance of the Perihelion of Mercury, which was an accepted fact among
astronomers, was puzzling since he knew that GTR is illogical. He knew, as did
all the Cambridge \ mathematical physicists, that Maxwell's Theory of
Electromagnetism was invariant with respect to Lorentz Transformations.
Law(iv) is probably the simplest Lorentz invariant modification of Newton's
\ Inverse Square \ law for gravity that could be devised.

It is \ clear from CN Ch. 8 that he had in mind the basic structure of his
final theory by late 1919 or early 1920 amd knew that it gave exactly the same
\textit{formula} for the Addvance of the perihlion as \ Einstein claimed for
GTR. He immediately set to work to find a theory consistent with his basic
Physical ideas ,set forth in PNK, CN and Part I of PRel and consistent with
observations. In fact \ he found FOUR logically valid theories, invariant with
respect to Lorentz ransformations, and Theories(i) and (iv) predict the
observved values for the perihelion and the bending of light. The "simplest"
of these is what he calls Law (iv) and is incorporated in what I am l calling
Theory(iv) which also encompasses Electromagnetic theory. In PRel he developed
(iv) whch is the "simplest" because no differential equations have to be
solved. The concept "impetus" plays a key role in enabling ANW to unify
gravitation and Electromagnetism in one theory. I am not certain whether he
checked that Theories (ii) and (iii) prredict correct values for the two
"classical" observations.

\ 

Part II sets out in considerable detail concrete physical consequences of
Theory(iv). Throughout his mathematical arguments ANW\ displays extraordinary
control of Lagrangian physics, inventiveness, and dligence.

Part III \ can be described \ quickly. It is a straight-forward exposition of
Tensor Calculus displaying it as a simple algebraic device for discussing a
mathematical system which is invariant\ under the action of a specified group
of linear transformations. For example: an orthogonal, a Lorentz, a Symplectic
or the full linear group \ of GTR.

ANW felt this was needed since most physicists in 1922 knew nothing about
Tensors and even were frightened by the thought of having to learn Tensor
Calculus in order to understand Einstein's theory which, in 1920, seemed
esoteric and impenetrable to most physicists. There is an amusing allusion to
this on p.182 of CN.

\bigskip

\subsection{\qquad IV. The criticism of \ Einstein's theory by ANW}

This has already been mentioned and is expressed in various contexts in
PRel\ and CN. It is essentially as follows:

- {\Large by identifying Whitehead's\textit{\ potential mass impetus}, }%
$dJ${\Large , such that }$dJ^{2}=J_{ij}dx^{i}dx^{j}${\Large \ for an interval
}$dx^{i}${\Large \ of the world-line of a particle, with Einstein's
\textit{distance} }$ds${\Large \ such that }$ds^{2}=g_{ij}dx^{i}dx^{j}%
${\Large , \ "Einstein leaves the whole antecedent theory of measurement in
confusion" (PRel, p.83). }

Einstein bases GTR on a vicious circle rather like "which came first, the
chicken or the egg?". The confusion can be illustrated in different ways. For example:

\textbf{A}. \ \ You cannot solve Einstein's equations, $G_{ij}=0,$ for
$g_{ij}$ until you enter the initial conditions and \textit{specify your
choice of units.} But you cannot explain your choice of unit of length until
you solve the equations and give a precise meaning to $ds^{2}$ \ because you
began by assuming arbitrary coordinates without defined units.

\textbf{B}. \ The idea of \ curved space and measurement in GTR is
often$^{13}$ glossed over with the charming story of The Student, The Ant and
The Apple. The 2-dimensional ant walks in a straight line from A to B on the
surface of the Apple, then to C and finally to A. He counts and records the
number of ant-steps in each side of the triangle ABC. \ He repeats this for a
continuous infinity of triangles and publishes a "Map of my Vicinity Recorded
in Units of \textit{Ant-steps}". \ It was generally agreed that "straight
line" should be understood to mean "geodesic". \ To carry out his program, the
ant would need an intuitive ability of recognizing the direction of the
tangent to the geodesic at of its each points , in other words of solving the
equations of a geodesic. \ Since no ant is known with this ability nor are
"nt-steps" a useful unit we agree that this gargantuan effort of the ant could
give no evidence about the curvature, if any, of the apple surface.\bigskip

Nor do I know any astronomer who claims to have the ability at any point in ST
of recognizing the direction of \ the tangent to the geodesic through the
given point to every other fixed point in ST. So, for example he cannot tell
how far the earth is from the sun.

\ 

This is such a simple clear vicious circle I can see no way out of it. \ Yet I
have never seen it directly addressed by the GTR enthusiasts among whom I must
place my younger self! \ Hitherto I have been a coward unwilling to contradict
what $everyone$ "knows" is firmly established. Also, until recently, I had not
understood how we were misled by Eddington.

\subsection{}

\subsection{\bigskip V. \ Eddington's Theorem}

\qquad In 1924, one year after his famous treatise on

General Relativity appeared, Eddington published a one-page note in NATURE
v.113, proving that, for a spherical static field surrounding a single
particle of mass $m,$ the metrical field $g_{ij}$ specified by
the\ Schwarzschild solution of Einstein's equations is isomorphic to the
gravitational field $J_{ij}$ described by ANW's Theory(iv) \textit{except }on
the variety $\ 2m=r.$

\bigskip

\textit{This theorem shows that Einstein's formula for the Advance of the
perihelion of Mercury can be justified by ANW's consistent Theory(iv) and that
curved ST is not needed to do so! \bigskip}

The proof of this key result can be found of course in NATURE, but also, more
relevantly$^{10}$ for what follows, in the\ second of the three Lectures by
Synge, in\ 1951 at the University of Maryland. These lectures are so valuable
that I made \ them available in arXiv, Physics/0505027. The main content of
that paper to which I refer as "JLS" is Synge's elegant summary of much of the
mathematical core of Part II of PRel. It also contains an Introduction by
myself \ in which is embedded a precursor of the present article, and an
Appendix, concerning the Solar Limb Effect, discussed briefly below which is
appended to this article.

It also suggests that any result claimed by GTR can be justified by PRel
without assuming curved ST if the Scwharzschild field is the only aspect of
GTR, beyond Newton, reqired in its alleged "proof". Possibly the results of
Probe B will be such a case.

\qquad\ 

The results of Probe B should be compared with predictions of Theory(iv) and
the e alleged conclusions of GTR.

\subsection{\qquad VI. Solar Spectral Shift}

\qquad This has a long and complex history. In 1907 what came to be named
\textit{The Limb Effect }was\textit{ }noticed\textit{ }\ Because of
cancellation of Doppler effect at the ends of an equatorial diameter of the
sun's disc we expect the average of frequency of the same line at the two ends
to equal that of the line at the center of the disc. The Swedish astronomer,
Dr. J.K.E. Halm$^{14}$, working in South Africa, was surprised that he found
very few lines which satisfied this expectation. The difference between the
expected and observed frequencies from line to line appears almost random.

According to my reading of the literature, this has had no satisfactory
explanation after more than a century.

In PRel \ Chs.10,11, 13-16, Whitehead presents his theory of the gravo-electro
interaction which does predict a Limb Effect! In 1946 when I still considered
GTR might be a reasonable theory, I drafted a paper examing the consequences
of his theory in the hope that it would resolve the advantages of Einstein's
versus Whitead's theories. It was accepted by Evershed, Editor of the
Astrophysical Journal, but because of change of jobs I did not mamage to
submit it in what I considered decent shape. Later in revised form it was
accepted for a Russian \ journal but the Journal never appeared! \ That paper
now appears here as the Appendix.

\qquad I consider it very important that this be updated by detailed study of
multiplets along the lines initiated by Miss Adam which may be the key to
unravelling the complexity of the Solar Spectrum.

This is imprtamt for our understanding of both movements of gas in the solar
atmosphere and also for the interpretation of the spectra of extra-galatic
stars and thus for all our cosmological speculations.

\subsection{%
%TCIMACRO{\TEXTsymbol{\backslash}}%
%BeginExpansion
$\backslash$%
%EndExpansion
}

\subsection{VII. \ Some Key\ Personae in the Drama\bigskip}

\textit{In this Section I assemble a few relevant considerations for which
there seemed no natural point-of-entry above.}

1. \textbf{John Keats}, the poet.

\qquad"Beauty is Truth and Truth is Beauty,

\qquad That is all ye know on Earth and all ye need to know."

This famous profession of faith of Keats is especially attractive to young
mathematicians when they realize how wonderfully simple GTR actually is.
Certainly that is how I felt.

We believed that since GTR is so simple it must be true! Perhaps it was this
that prevented us and many others from seeing the obvious validity of ANW's
Criticism of GTR.

2 \textbf{Eddington}

\qquad It is not widely realized that the existence of Dirac's equation for
Hydrogen was a big shock for A.S.Eddington(ASE). \ It shattered his firm
belief that Tensor Calculus is totally sufficient for discussion of
Relativity. This led him to develop an elaborate theory about \textit{The
Constants of Nature }by\textit{ }which he predicted the values of ten
constants. By a method not available to Einstein, ASE used three of these to
establish Units and obtained all the remaining with remarkable accuracy. No
physical theory has claimed such wide-ranging objective confirmation. But it
was dismissed as nonsense by most physicists of the thirties as
\textit{philosophical speculation }by a man in his dotage!! \ For example such
was the attitude of Heisenberg with whom I spent two evenings in his apartment
in Goettingen in 1946.

Eddington fixed on the fact that the four matrices, used by Dirac to formulate
his famous equation, generate a 16-dimensional Algebra$^{15}$ over the complex
numbers which we now call a Clifford Algebra.\ \ ASE attributed such a
16-dimensional algebra to each electron and each proton to carry its
\textit{internal structure w}hich may thus be considered a major
generalization of "spin"\textit{.}\ The theory was developed in a long series
of papers which were summarized in two$^{16}$ major monographs in 1936 and,
posthumously, in 1947.

\textbf{3. \ Synge and Schroedinger}

\qquad In 1952 Synge invited me to give eight lectures to his Seminar in
Dublin with himself and Schroedinger, by no means silent, in the front row!
\ These were about my 1943 Toronto Thesis based on ASE's Theory of the
Constants of Nature for which Synge had been one of my Examiners. My
Supervisor was Einstein's coauthor, Leopold Infeld. \ Unlike Heisenberg and
other physicists with whom I talked neither Schroedinger nor Synge were
willing to dismiss out-of-hand the last fourteen years of work of \ the man
who essentially created modern Astrophysics with his book$^{17}$ on the
\textit{Internal Constitution of the Stars.}

Personally, though\ there were a few minor and one major mathematical errors
in his argument, I gained great respect for Eddington's physical insight and
for his desire to stick to physical fact. \ I have never noticed these virtues
in writngs about String Theory and related theoretical activity most of which
I regard as mathematical romanticizing.

4\textbf{. Will's Important Paper}

\ \ \ \ \ \ In 1971, Clifford M. Will$^{18}$ published a paper which has
played a very significant role in determining how the physics community dealt
with Whitehead's Theory (iv). This paper was written before Will had finished
work on his Ph.D. under the Supervision \ of Kip Thorne. At the same time he
was working on several papers with Thorne and\ others. He implies that he had
not read Whitehead in any depth. This indeed was confirmed in my mind by his
Footnote (10) in which he ascribes to ANW the opposite of my understanding of
Whitehead's choice of units. \ He concluded that for a certain period in the
analysis of the motion of tides Theory (iv) predicts a result which is in
error by 200 times the estimated possible error and therefore was sure that he
had delivered the \textit{coup de grace }to ANW's Theory. This conclusion was
accepted in the well-known comprehensive treatise, GRAVITATION, by Charles
Misner, Kip Thorne, and \ John Wheeler (MTW) and/ so became part of our
accepted Dogma.

\qquad In the only critical discussion of Will's paper I have seen
Fowler$^{19}$ replaces a basic assumption of Will by an alternative which
seems reasonable and concluded that Will had over-estimated ANW's error by a
factor of 100! I drew Fowler's paper to Clifford Will's attention some years
ago and also pointed out that Dark Matter had not been thought of when he and
Fowler wrote and that the existence if Dark Matter would change his
conclusion. He felt that this had only an insignificant effect on his
argument. I also asked him to rebutt Whitehead's Criticism of GTR, but with no success.

More recently$^{20}$ Gary Gibbons and Cliford Will have issued a paper with an
improved version of the previous paper. They announce four new criticisms of
Theory (iv).

\qquad I infer from p.84 of PRel that ANW suspected that his Theory (iv) was
not Final. This is partly why he offered three alternative possibilities for a
gravitational LAW. Having found LAW (iv) which was Lorentz invariant,
logically consistent and predicted the observed values for he two bits of
\ evidence on which Eddington had based his advocacy of GTR, he turned his
attention to matters of greater interest to himself. Knowing that he had
disposed of GTR as a viable theory Whitehead never thereafter referred to it
positively as far as I have noticed .

\subsection{\bigskip}

VIII. \textbf{Concluding Observations}\bigskip

\textit{Anyone acquainted with current cosmolgical speculation will realize
that the implications of my thesis are almost devastating. Since I do not
claim to have mastered String Theory or the Standard Model I appeal to many
who are more competent than I\ to work out in detail the consequences of this
paper. }

1. \textbf{Humility}

\qquad Perhaps the most important is the realization that humanoids\ must
humbly accept the fact that we cannot formulate in human langauge a "Final
Theory". The presumption that he could is essentially why Adam was thrown out
of Eden! \ Personally I have always preferred Whitehead's step-by-step
approach to unsderstanding physics than Einstein's exaggerated hopes and claims.

2. \textbf{Theory (iv) is currently the best}

\qquad At present there is no valid evidence that ST is curved but rather the
opposite. \ As Whitehead noticed betweem 1917 and 1920, \textit{any theory of
gravitation which makes the metric of Space-Time dependent on the
circumambient masses is not viable. }

Einstein became famous and GTR thereafter dominated our thinking because of
the claim that GTR predicted the two Classical Tests. In fact, the observed
results were predicted not by GTR but what I now call the "mongrel" version of
GTR which implicitly assumes that ST is flat. ANW's consistent Theory (iv)
explicitly makes the same predictions. In my lectures about GTR, following
Eddington and most textbooks I justified the mongrel theory by arguing that
locally the gravitational fields are so weak that the Einsteinian and
Newtonian metrics could not be observationally distinguished. \ This may be
true "observationally " currently but there is a distinction or the earth
would not move in an ellipse! But this justification fails completely for the
strong fields which increasingly are of cosmological interest.

In my opinion, Theory (iv) will probably be easier to reconcile with QM and it
is not based on a vicious circle. So currently it is the best available theory
even though, thanks to Clifford Will, we know that as Whitehead expected it is
not Final.

Eddington in connection with his Theory of the Constants of Nature, faced and
solved the problem of units. His was a wide-ranging Theory which predicted
more than ten quantities. By chosing three deternmined by functionally
independent formulas and setting them equal to accurately observed values he
was able to establish meaningful units of length, time and mass.\ \ So
equipped, he could predict seven Constants with extraordinary accuracy.

. \ It seems to me that at present the most useful expenditure of time and
mental energy would be to tackle the problem of reconciling ANW's Theory (iv)
with QM. Possibly a first step would be to pursue the problem introduced in
the Appendix. Or to consider whether ANW's formula for the gravo-elctro
Impetus can be modified with advantage by ideas from Quantum Field Theory
and/or ANW's metaphysics.

3 \textbf{Extend ANW's Unified \ Theory}

\qquad PRel already includes gravitational and electromagnetic forces. in the
definiton of Impetus Potential in Equation (9) .p. 79. Can we obtain analogous
terms for Weak and Strong forces? This would provide a Unified theory of
forces now known.

4 \textbf{Important Research Problem}

\qquad It could be of considerable significance if the research begun in the
Appendix were checked and completed. \ This would entail a detailed study of
the variations of frequency of lines of multiplets in the solar spectrum. As
observed by Miss Adam$^{21}$ of the Oxford Observatory, the Limb Effect for
different lines in the muultiplet vary in an astonishing manner. \ It would be
desirable to verify, extend and explain her observations.

It seems to me that the observed differences are caused only by gravitational
or possibly pressure effects. That is, they provide a considerable body of
data for which it might be relatively easy to obtain a convincing explanation.
\ Armed with this we might be able to understand the total spectrum with more
assurance than presently is the case.

This could help us to interpret the motion of gases in the Sun's atmosphere
and also the spectra of distant stars on which now much cosmological
speculation rests.

5. \textbf{Back to Eddington?}

\qquad It may be worthwhile to return to Eddington's Theory, seeking to
understand how he obtained such remarkable predictions

6. \textbf{Revisit Black Holes with ANW?}

\qquad It may well be very interesting to develop the theory of Black Holes
using ANW's theory as set forth by Synge in the Proc.of the Royal
Society$^{22}$ and in the last few pages of his Maryland Lectures$^{10}$.
\ ANW's treatment of the field of an isolated\ massive point-particle does not
have the Schwarzschild singularity.

\subsection{\bigskip IX. \ \ \ \ APPENDIX}

\textbf{WHITEHEAD' S PERTURBATION OF ATOMIC ENERGY LEVELS}

\ \ \ \ \ \ \ \qquad\qquad\qquad\qquad\ \ \ A. J. Coleman

Department of Mathematics, Queen's University,

Kingston, Ontario, Canada.

\bigskip dolemana@post.queensu.ca

Whitehead's theory of relativity implies that there is an interaction between
the gravitational and electromagnetic fields such that for an atom at the
surface of a star, the Coulomb potential r$^{-1}$between two charges must be
replaced by%

\begin{equation}
\frac{1}{r}(1-\alpha cos^{2}\theta).\ \ \tag{$\left(                 1\right)
$}%
\end{equation}
\ \ \ \ \ \ \ \ \ \ \ \ \ \ \ \ \ \qquad\qquad\qquad\ \ \ \ \ \ \ \ 

Here, $\theta$ is the angle between the radius vector\ joining the two
interacting charges and the \ direction of the stellar radius passing through
the atom; $\alpha$ is a constant depending on the strength of the
gravitational field. \ At the surface of the sun, $\alpha$ = 2.12$\times
$10$^{-6}$ approximately.

The effect of (1) is to perturb the normal energy levels by the small term
\qquad%

\begin{equation}
-\alpha\frac{\cos^{2}\theta}{r} \tag{$\left(           2\right)            $}%
\end{equation}
which has axial symmetry about the stellar radius through the atom. \ An
effect of precisely this symmetry is what is needed to explain the limb-effect
in the solar spectrum. \ One might also hope that this perturbation could
account for the striking differences which have been observed in shifts within
the same solar multiplet.

The effect of the perturbation (2) acting between all pairs of charge is to
add
\begin{equation}
V^{`}=\Sigma_{i}\alpha\frac{Ze^{2}}{r}\cos^{2}\theta_{i}-\Sigma_{i<l}%
\frac{\alpha e^{2}\cos^{2}\theta_{ij}}{r_{ij}} \tag{$\left(      3\right)  $}%
\end{equation}
to the potential in Schroedinger's equation. Here, Ze is the charge of the
nucleus; 1%
%TCIMACRO{\TEXTsymbol{<} }%
%BeginExpansion
$<$
%EndExpansion
i, j%
%TCIMACRO{\TEXTsymbol{<} }%
%BeginExpansion
$<$
%EndExpansion
N , where N is the number of electrons in the atom; $\theta_{i}$ is the angle
between \textbf{r}$_{i}$ and the\textquotedblleft vertical\textquotedblright%
\ ; $\theta_{ij}$ is the angle between \textbf{r}$_{ij}$ and the vertical.

By first-order perturbation theory, the shift in energy of a J,M level is
\begin{equation}
\Delta E_{JM}=\langle JM|V^{`}|JM\rangle\tag{$\left(                 4\right)
$}%
\end{equation}

For a Term with total orbital momentum L and spin S ,
\begin{equation}
|JM\rangle=\Sigma_{\mu+\nu=M}\langle L\mu S\nu|JM\rangle\varphi_{L\mu}U_{S\nu}
\tag{$\left(                                                   5\right)  $}%
\end{equation}
where $\langle$L$\mu$ S$\nu$%
%TCIMACRO{\TEXTsymbol{\vert}}%
%BeginExpansion
$\vert$%
%EndExpansion
JM$\rangle$ is the vector coupling coefficient, and $\varphi_{L\mu}$ and
$U_{S\nu}$ are, respectively, the appropriate pure orbital and pure spin
functions. Since
\begin{equation}
ccs^{2}\vartheta=\frac{1}{3}+\frac{1}{3}(3\cos^{2}\vartheta-1) \tag{$\left(
6\right)                                                   $}%
\end{equation}
by using the indistinguishability of the electrons, the perturbation (4) can
be expressed in the form%

\begin{equation}
\Delta E_{JM}=<JM||V_{0}|JM>+<JM|V_{2}|JM>, \tag{$\left(            7\right)
$}%
\end{equation}
\newline where
\begin{equation}
V_{0}=\frac{\alpha}{3}N\left(  \frac{Ze^{2}}{r_{1}}-\frac{N-1}{r_{12}}%
e^{2}\right)  ,\quad and \tag{$\left(                   8\right)    $}%
\end{equation}

\bigskip%

\[
V_{2}=\frac{\alpha}{3}Ne^{2}\left[  \frac{Z}{r_{1}}\left(  3\cos^{2}%
\vartheta_{1}-1\right)  -\frac{N-1}{2r_{12}}\left(  3\cos^{2}\vartheta
_{12}-1\right)  \right]
\]

$\qquad\qquad\qquad\qquad\qquad\qquad\qquad\qquad\qquad\qquad\qquad
\qquad\qquad\left(  9\right)  $

\qquad\qquad\qquad\qquad\qquad

The advantage of this decomposition is that with respect to simultaneous
rotation of all electrons about the nucleus, V$_{0}$ and V$_{2}$ belong to
D$_{0}$ and D$_{2}$ \ \ representation of the rotation group, respectively.

An application of the Wigner-Eckhart theorem leads to the conclusion that
\begin{equation}
\Delta E_{JM}=A_{L}+\frac{3M^{2}-J\left(  J-1\right)  }{J\left(  2J-1\right)
}B_{J} \tag{$\left(  10\right)   $}%
\end{equation}

where
\begin{equation}
A_{L}=\langle\varphi_{LL}|V_{0}|\varphi_{LL}\rangle,\quad B_{J}=\langle
JJ|V_{2}|JJ\rangle. \tag{$\left(               11\right)                $}%
\end{equation}

By employing (5) and the theory of vector-coupling coefficients, a rather
tedious calculation results in the formula%

\begin{equation}
B_{J}=\Sigma_{\mu+\nu=J}\left|  \langle L\mu S\nu|JJ\rangle\right|  ^{2}%
\frac{3\mu^{2}-L\left(  L+1\right)  }{L\left(  2L-1\right)  }B_{L}=
\tag{$\left(                                                   12\right)   $}%
\end{equation}

\qquad\bigskip$\{1+\frac{3(J-L-S)(J+S-L+1)[(J+L-S+1)(J+L+S+2).-2J-3]}{L\left(
2L-1\right)  \left(  2J+2\right)  \left(  2J+3\right)  }\}B_{L}$

\bigskip

where
\begin{equation}
B_{L}=\langle\varphi_{LL}|V_{2}|\varphi_{LL}\rangle. \tag{$\left(   13\right)
$}%
\end{equation}

It follows from the Virial Theorem that%

\begin{equation}
A_{L}=-2\alpha E_{L} \tag{$\left(                14\right)                 $}%
\end{equation}

where E$_{:L}$ is the total energy of the state $\varphi_{LL}$ which is given
with sufficient accuracy for the present purposes by the mean observed energy
of the Term. Thus for a Fraunhofer line, the V$_{0}$ tern gives rise to a
red-shift which is proportional to the wave\-length of the line and equal to
2/3 of that predic$t$ed by Einstein.

We have thus reduced the problem of calculating the Whitehead shift in the
levels of a Term to that of evaluating the one constant B$_{L}$. Consequently,
the shifts in the lines of a multiplet depend on two constants B$^{i}$ ,
B$^{f}$ associated with initial and final levels.

Dr. R M. Erdahl has suggested that in attempting to check this theory against
observations we should treat B$^{i}$ and B$^{f}$\ as phenomenological
constants. In certain cases B$_{J}$ = 0 , so that for these the predictions
are particularly simple. For example, from (13) it follows immediately that
B$_{L}$ = 0 if \ L = 0, that is for $an$ S-term. But it follows from (12) that
B$_{J}$ also vanishes for states such as $^{4}$P$_{1/2}$ , $^{6}$%
D$_{1^{\prime}2},$ ... $^{10}$F$_{11/2}$ and many others. \ It may also be
worth looking at Terns for which B$_{J}$ is small. \ 

To test the usefulness of Whitehead's perturbation in explaining the actual
complex observations of shifts in the solar s$p$ectrum, it would be
particularly valuable to have reliable measurements for the absolute shifts at
various points in the solar disc for all lines of a multiplet and especially
for multiplets which include one or more transitions between energy levels
with symmetry type appearing in the list \ described above.

In addition to possible perturbation of energy levels by a gravo-electric
interaction, the Fraunhofer lines are undoubtedly shifted by Doppler and
pressure effects. To this must be added the classic Einstein shift which has
been confirmed by the Pound-Rebka experiment and which follows from Newton's
theory and the conservation of energy. The Einstein and Doppler shifts are
proportional to the wave-length of the line and by themselves certainly cannot
explain the observed shifts in the solar spectrum.

If Whitehead's perturbation combined with reason\-able assumptions about
pressure shift is unable to ex\-plain the observations, all is not lost. If
the astron\-omers can obtain reliable observations, especially at the limb$, $
of a large number of multiplets of diverse symmetry, it should be possible,
using the techniques of the present paper, to obtain a good approximation for
a perturbation of atomic energy levels which would explain the observations by
employing a multipole analysi$s$.

Since in the solar spectrum, the observed deviations from Einstein's predicted
shift are as large or larger than his prediction, it is clearly of great
interest to establish the source of this deviation in order to be able to
interpret spectral shifts from other stars with any confidence.

\bigskip

August 16, 1968.

\bigskip

(\textbf{NOTE}. \ This paper had a really strange history! In a letter dated
September 10,1948, a first draft was reommended to J.E. Evershed, subject to
revisions, for publication in the Astrophysical Jounal by H.H. Plaskett,
Director of the Oxford Observatory. Though I presented its ideas at a couple
of conferences, because of changing professional duties only twenty years
later was it revised to its above form.

For many years, I assumed that this paper had appeared in the Proceedings of
the International Conference on Relativity and Gravitation in the USSR which
R. M. Erdahl and I attended in 1968 and where it was delivered and accepted.
\ Only in 2003, when my old interest in Whitehead's theory was reviving, did
my friend Prof. V. I. Yukalov inform me \ that the Proceedings of the
Conference were \ never published. AJC ).

\bigskip

\subsection{\bigskip X. Notes}

1. \ Trilogy, Cambridge University Press

\ 

\qquad\ \ 1919 \ An Enquiry Concerning the Principles of Natural

\ \ \ \ \ \ \ \ \ \ \ \ \ \ \ \ Knowledge (PNK).

\qquad\ \ 1920 \ The Concept of Nature(CN).

\ \ \ \ \ \ \ \ 1922 \ The Principle of Relativity(PRel).

2. \ Albert Einstein, Zur allgemeinen Relativitaetstheorie,(On the\ General
Theory of Relativity), 1915, Prussian Academy of Sciences

3. In Lowe's biography of ANW$^{7}$ the reader can find bibliographical
details of Whitehead's published works, pp.367/373 of Volume II, and in Vols.
I\&II background information for the following selection of items:

(i) 1905 was the year in which ANW\ submitted \textit{On Mathematical Concepts
of the Material World }\ to the Ph.Trans.of the RS,London,Ser.A,205(1906) 465-525

(ii) \ \textit{Space,Time and Relativity, }Proc.Aristotelian Soc., N.S. 16
(1915-1916): 104-29

(iv)\textit{Einstein's Theory: An Alternative Suggestion} Times(London)
Ed.Supp. Feb. 12,1920, p. 83. This brief note together with Ch.8 of CN
convince me that ANW had realized by the end of 1919 that GTR was
\textit{logically }non-viable and that he had the main contents of PRel in his
head!\ \textit{ }

(v-vii) The Trilogy, cf. Note(1)

(viii) \textit{Science and the Modern World}, Macmillan,1925

(ix) \ \textit{Process and Reality -An Essay in Cosmology, Macmillan, 1929.}

4. \ H. Keeton makes the same point in the book edited by Timothy Eastman and
himself: \textit{ PHYSICS and WHITEHEAD - Quantum Process and Experience,
}2004\textit{, }State University of New York. ISBN 0-7914-5913-6
+paperback\textit{.}

5. \ Reinhold Niebuhr was one of the most important American protestant
theologians. His attitude to the task of the Christian Church in the USA was
basically changed in the 1930's by the suffering of members of the
working-class in his Parish and the apparent irrelevancy of the Church's
message to them. He called for a radical "trans-valuation\ of values".
Scientists like myself often rightly excorciate so-called "fundamentatlists"
for being too slow in absorbing Niebuhr's message but we are as bad or worse
in showing little signs that we have absorbed ANW's wisdom of almost a century ago!

6. \ As a teen-ager I ate up Eddington's marvelously clear expositions of GTR
and QM. So when in a second-hand book-store, between my junior and senior
years undergraduate Course \ in Toronto U., I saw a book for \$2.50 called
\textit{The Principle of Relativity. }I picked it up eagerly assuming it was
about GTR!. While Whitehead, the author, praised Einstein to the skies I was
surprised that he pointed to a\ logical flaw in GTR. His criticism has haunted
me therefore since I was 20 years old!

\textbf{7}. Victor Lowe's two-volume biography, \textit{ALFRED NORTH WHITEHEAD
- The Man and His Work, }1985 \& 1990, The Johns Hopkins Universtiy Press, is
regarded as definitive. It certainly provides a valuable detailed
Bibliography\ which I found very useful.

8. \ I discovered that a vivid, dramatic and amusing portrayal of the
intensity of the Einstein euphoria results from skimming through on the Web
the first-hand accounts in the NY Times of Einstein's visit to the USA in 1920.

9. "disdain" is not too strong a word for the feeling that I often sensed
among hard-nosed physicists who for several decades seemed to regard
themselves as the only percipients of Truth. Though, fortunately, the
following attitude is not common, it contains such an expressive turn of
phrase I put it on record. An Australian friend of mine obtained a Ph.D. in
Life Sciences. At one point he sought advice from his Supervisor. \ "Would it
be worthwhile for me to study the Philosophy of Organism propounded by the
philosopher Whitehead?"... "For Heaven's sake, NO! You are surely old enough
to realize that Philosophy is nothing but mental masturbation." \ 

10. \ The lectures of J.L. Synge on Dynamics, Tensors and GTR were the most
precise and elegant mathematical lectures I ever attended. His two-volume
treatise on Relativity: Special, 1955, General,1960 was characterized by an
astrophysicist friend as the best currently available. In 1952 when he invited
me to lecture to his Seminar at the Dublin Institute we became and remained
friends until his death in 1995. Even so I regret that his failure to
understand Part I of PRel and his penchant for geometrizing all of physics
helped to prevent us from grasping the truth of PRel. In fact he came close to
seeing the point with his remark, p.296/7 of his book on GTR, that Einstein's
method of proving the validity of his prediction of the Advance of the
perihelion of Mercury was "intellectually repellent"! \ This came close to
saying that the prediction was made by a "mongrel" theory.

11. Dean R. Fowler presented a Ph.D. thesis in Theology of which the first
part contains an insightful account of ANW's approach to Physics. In writing
this he had assistance from an competent theoretical Physicist.\ This is
summarized in \textit{Process Studies, }5, 159-174$\left(  1975\right)
.$Based on this he criticized \ Will's article, asserting that its estimate of
ANW's error was 100 times too large, ibid. 4,4(1974). \ More recently , the
philosopher Gary Herstein, has published a book, setting forth in some detail
the criticism of GTR by ANW: \ Herstein, Gary L. \textit{Whitehead and the
Measurement Problem of Cosmology}, Ontos-Verlag, Frankfurt / Lancaster / New
Brunswick, Process Thought V, June 2006.

12. In section 42 of the \textit{Dialogues of Alfred North Whitehead , }by
Lucien Price, of which there have been several editions including paperback,
Whitehead states in forceful language his conviction that humanoids will never
be able to fashion a "final" theory.

13. \ This cute story is employed, for example, in the widely used textbook
\ GRAVITATION \ by Misner,Thorne and Wheeler.

14. \ The discovery of the Limb Effect is attributed to the Swedish
astronomer, J.K.E. Halm,who was Chief Assistant at the Cape Observatory in
South Africa from 1907 to 1927. It is believed that in 1907 he and the
Astronomer, Hough were observing the frequencies of lines from various points
on the solar disc.

15. In analogy with Dirac's equation, to a basis (e$_{i}$) of 4-dimensional
Euclidean space, Eddington associated four 4 $\times$4 matrices (E$_{i}$) such
that%
\[
E_{i}E_{j}+E_{j}E_{i}=2\delta_{ij}I
\]
where $I$ is the identity operator. By taking all possible products of the
$E_{i}$ then by using the above anticommutation properties we find that there
are 16 linearly independent products which span a 16-dimensional linear space
over the complex numbers. This space can be regarded as a direct sum of the
6-dimensional subspace of of skew-symmetric matrices and the 10-dimensional
space of symmetric matrices. ASE attributed the electromagnetic properties of
the particle to the first of these subspces of dimension 6 and the mechanical
properties to the symmetric sub-space. Analogously the internal properties of
two particles requires a linear space of dimemsion 16$\times$16=256 dimensions
with two subspaces of dimension of 120 and 136. By more than one argument, the
first in 1931, ASE was led to consider the equation%
\[
10m^{2}-136mm_{0}+m_{0}^{2}=0
\]
with roots of ratio \ 1847.60 which is so close to the ratio of mass of proton
to electron that ASE, the physicist, could not ignore it. \ Until his death in
November 1944 his intense research effort was devoted to understanding this
and several other numbers to which he was led.

16. \ \textit{Relativity TTheory of Protons and Electrons}, Cambridge
University Press, 1936 (RTPE); \ \textit{Fundamental Theory}, ibid, 1946 (FT)

Interested readers might begin with the Preface of FT and the Table on p.66.

17. T\textit{he Internal Constitution of the Stars}, Cambridge University
Press, 1930.

18. Clifford M.Will, Astrophysical Journal, 169,141-155(1971)

19. \ \textit{Process Studies}, 4,4

20. \ Gary Gibbons and C.M. Will, \textit{arXiv:gr-qc/0611006v1}\ 1 Nov 2006.

21. \ Madge Adam, Mon.Notices, RAS,119,460-470(1959); ibid.177,687-707(1976).
In these and other papers she reports shifts for a variety of multiplets.

22. \ J. L. Synge, \ Proc. Roy. Soc., London\ A211(1952) 303.

\bigskip

\textit{Acknowledgements}. I am grateful to John Lindsay, Timothy Eastman and
Ronny Desmet for essential assistance during the productin of this paper.

\bigskip

Easter, 2007
\end{document}